\documentclass[submitting]{nst}
\usepackage{subfigure,dcolumn}
\usepackage{epstopdf}
\usepackage{mhchem}
\usepackage{upgreek}
\usepackage{float}
\usepackage{hyperref}
\usepackage{diagbox}

\begin{document}

\title{Classifier for centrality determination with Zero Degree Calorimeter at the Cooling-Storage-Ring External-target Experiment}
\thanks{This work was supported in part by the National Nature Science Foundation of China (NSFC) under Grant No. 11927901 and 12175084, the National Key Research and Development Program of China under Grant No. 2020YFE0202002 and 2022YFA1604900, and the Fundamental Research Funds for the Central Universities (No. CCNU22QN005).}

\author{Biao Zhang}
\affiliation{Key Laboratory of Quark \& Lepton Physics (MOE) and Institute of Particle Physics, 
Central China Normal University, Wuhan 430079, China}

\author{Li-Ke Liu}
\affiliation{Key Laboratory of Quark \& Lepton Physics (MOE) and Institute of Particle Physics, 
Central China Normal University, Wuhan 430079, China}

\author{Hua Pei}
\affiliation{Key Laboratory of Quark \& Lepton Physics (MOE) and Institute of Particle Physics, 
Central China Normal University, Wuhan 430079, China}

\author{Shusu Shi}
\affiliation{Key Laboratory of Quark \& Lepton Physics (MOE) and Institute of Particle Physics, 
Central China Normal University, Wuhan 430079, China}

\author{Nu Xu}
\affiliation{Key Laboratory of Quark \& Lepton Physics (MOE) and Institute of Particle Physics, 
Central China Normal University, Wuhan 430079, China}

\author{Yaping Wang}
\email[Corresponding author, ]{wangyaping@mail.ccnu.edu.cn}
\affiliation{Key Laboratory of Quark \& Lepton Physics (MOE) and Institute of Particle Physics, 
Central China Normal University, Wuhan 430079, China}


\begin{abstract}

The Zero Degree Calorimeter (ZDC) plays a crucial role in determining centrality at the Cooling-Storage-Ring External-target Experiment (CEE) in the Heavy Ion Research Facility in Lanzhou (HIRFL). A Boosted Decision Trees (BDT) multi-classification algorithm is employed to classify the centrality of the collision events based on the raw features from ZDC such as the number of fired channels and deposited energy. The data from simulated $\rm ^{238}U$ + $\rm ^{238}U$ collisions at 500 $\rm MeV/u$, generated by the IQMD event generator and subsequently modeled through the GEANT4 package, is employed to train and test the BDT model. The results showed the high accuracy of the multi-classification model adopted in ZDC for centrality determination, which is robust against variations in different factors of detector geometry and response. The study demonstrates a good performance of the CEE-ZDC for determining the centrality in nucleus-nucleus collisions.

\end{abstract}

\keywords{ZDC, Boosted Decision Trees, multi-classification, IQMD, centrality determination}

\maketitle
\nolinenumbers

\section{Introduction}\label{sec.I}
The primary objective of heavy-ion collisions at different beam energies is to investigate strong interaction matter and comprehend the QCD phase diagram. The phase diagram provides information on the phase transition and critical point of the strongly interacting system, where hadron gases exist at lower temperatures and low baryon density, while at higher temperatures or densities, the hadronic boundary disappears, and confined quarks move freely throughout the system~\cite{Braun-Munzinger:2007edi}. The Beam Energy Scan program of RHIC-STAR aims to approach the possible critical point from the high-energy side. Still, it is essential to study the phase diagram in the hadron phase and approach the critical point from the low-energy side~\cite{Bzdak:2019pkr, Huang:2023ibh, Luo:2020pef}. The Cooling-Storage-Ring External-target Experiment (CEE) at the Heavy Ion Research Facility in Lanzhou (HIRFL), with its advanced spectrometer, provides significant opportunities to study the phase diagram at extremely high net baryon density levels with energies of several hundred AMeV~\cite{CEE:T02019}.

The Zero Degree Calorimeter (ZDC), one of the sub-detector of CEE in the forward rapidity region, is designed to accurately determine the centrality and reaction plane of the collision events~\cite{CEE:ZDC2021}. The collision events are typically classified into centrality classes representing certain fractions of the total reaction cross sections corresponding to specific intervals of impact parameter $b$~\cite{Svetlichnyi:2021vox}. Impact parameter $b$ is an essential parameter to understand the initial overlap region of the colliding nuclei of collected data in heavy-ion collisions, which represents the distance between the nuclei centers in the plane transverse to the beam axis, determining the size and shape of the resulting medium. However, the impact parameter $b$ is not directly measurable in experiments. To estimate centrality experimentally, the raw observables that scale monotonically with impact parameters could be used for classification according to centrality, e.g. the reconstructed tracks with central-barrel tracking detectors or the deposited energy in the forward calorimeters. Accurate centrality determination is a baseline for many physics analyses in heavy-ion collision experiments~\cite{Altsybeev:2016uql}, particularly in the study of the search of observables sensitive to a possible phase transition or critical point by the analysis of fluctuations and correlations. 

In recent years, Machine Learning (ML) methods have gained significant attention for determining centrality class in heavy-ion collisions \cite{Altsybeev:2016uql, OmanaKuttan:2020brq}. Previous studies have treated centrality determination as a regression problem on impact parameters and utilized combined information from central tracking systems and forward calorimeters to train ML models. However, to avoid auto-correlation in physics analysis, this paper adopts a machine learning approach that solely utilizes raw experimental features from the forward calorimeter to determine centrality. We report the application of a multi-classification ML algorithm based on Boosted Decision Trees (BDT) as a centrality classifier using solely ZDC in $\rm ^{238}U$ + $\rm ^{238}U$ collisions at 500 $\rm MeV/u$ at the CEE. The ML inputs were generated using the Isospin dependent Quantum Molecular Dynamics (IQMD) generator \cite{Hartnack:1997ez}. Additionally, we present efficiency and purity measures related to the centrality determination performance of the ZDC with the model application.

\section{CEE-ZDC}\label{sec.II}

\begin{figure}[htb]
  \subfigure[] {
   \label{Fig:1a}     
  \includegraphics[width=0.48\columnwidth]{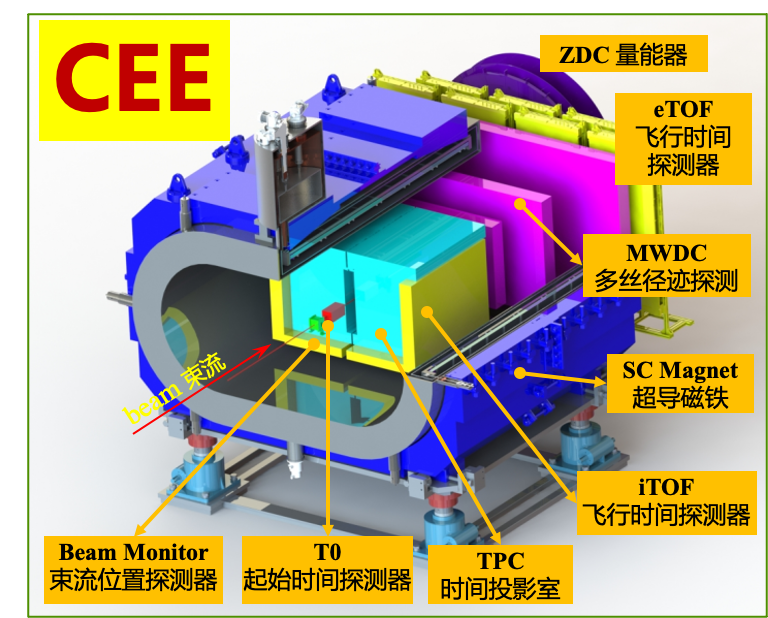}
  }
  \subfigure[] { 
  \label{Fig:1b}     
  \includegraphics[width=0.4\columnwidth]{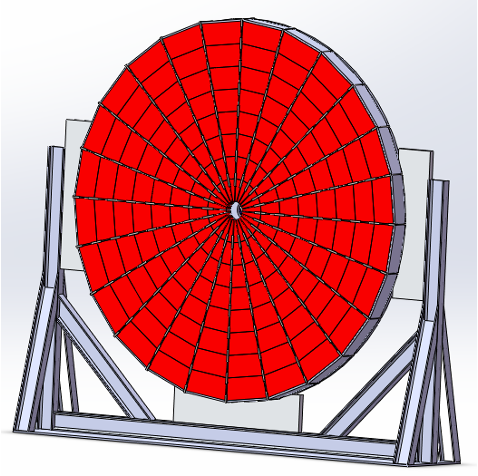}
  } 
  \caption{(a) CEE detectors schematic layout. (b)  ZDC detector layout.}     
  \label{Fig:CEE_ZDC}
\end{figure}

The CEE, which adopts fixed-target-mode heavy-ion collisions, is the first large-scale nuclear experimental device operating in the GeV energy region in China. It is equipped with a set of sub-detectors, as shown in Fig.~\ref{Fig:1a}. The CEE detector system comprises a beam monitor, T0 detector~\cite{CEE:T02019}, time projection chamber (TPC)~\cite{Huang:2018dus}, internal time-of-flight (iTOF) detector~\cite{CEE:iTOF2022}, a large superconducting dipole magnet, multiwire drift chamber (MWDC)~\cite{Sun:2018kmj}, external time-of-flight (eTOF) detector~\cite{CEE:eTOF2020}, and a zero-degree calorimeter (ZDC)~\cite{CEE:ZDC2021}.

The ZDC is centrally positioned at the end of the CEE, covering the pseudorapidity range of $1.8 < |\eta|< 4.8$. The ZDC utilizes a symmetrical and fan-shaped layout, with 8 radial and 24 angular sections, and a maximum radius of 1 meter. The detector comprises trapezoidal modules equipped with uniform plastic scintillators that are coupled with a light guide and then connected to photomultiplier tubes (PMT) to convert charged particles into charge signals. To obtain a comprehensive signal, each module produces two charge signals for two dynodes of each PMT that are transmitted to two separate readout channels, resulting in a total of 384 (24 × 8 × 2) channels for the ZDC. The purpose of ZDC is to detect particle fragments in the forward rapidity region following semi-central and peripheral collisions, providing vital information for the precise reconstruction of the centrality and reaction plane of collision events~\cite{CEE:ZDC2021, Liu:2023xhc}.

\section{Model training with simulated Event}\label{sec.III}

\begin{figure}[htb]
  \subfigure[] {
   \label{Fig:2a}     
  \includegraphics[width=0.46\columnwidth]{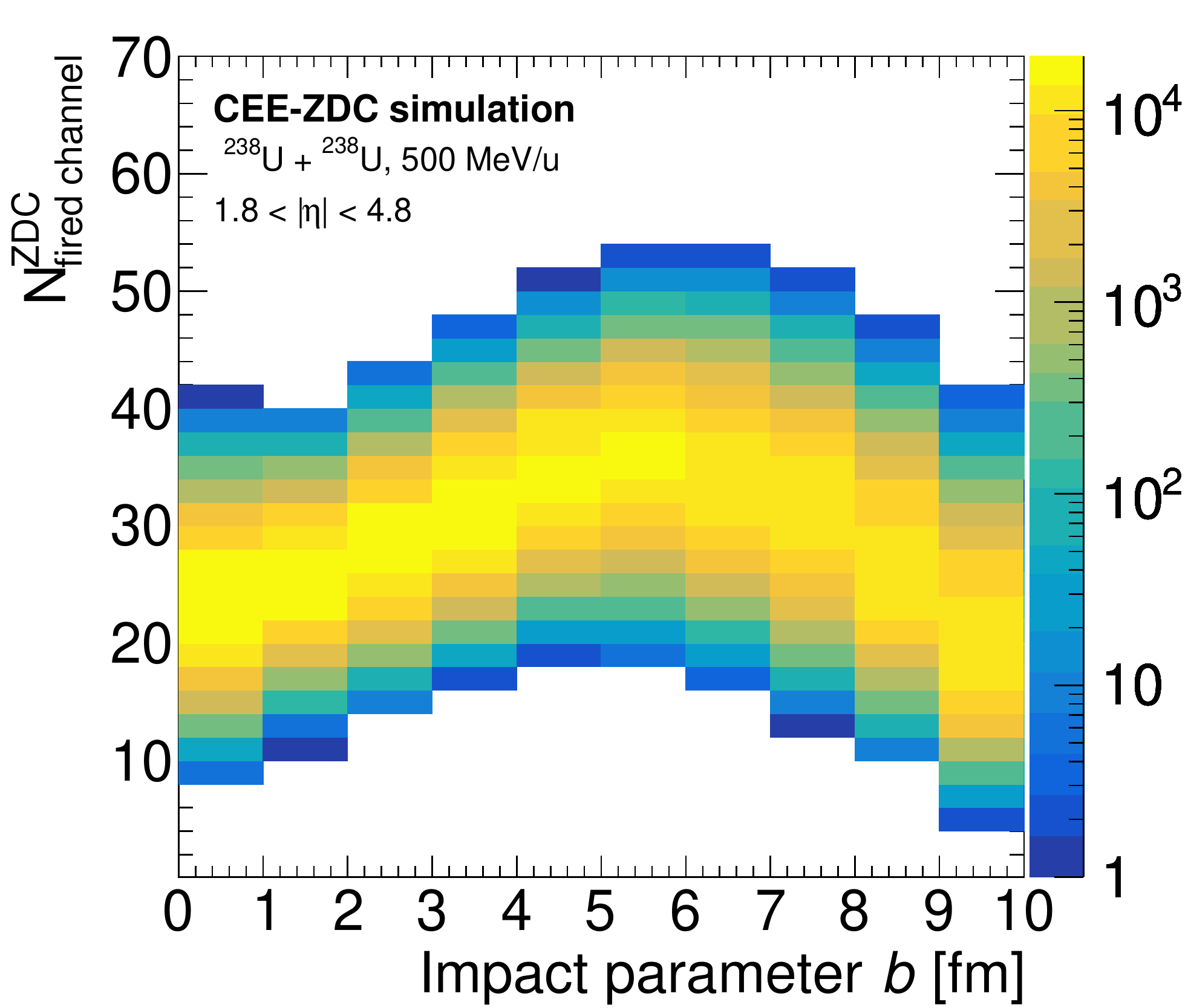}
  }
  \subfigure[] { 
  \label{Fig:2b}     
  \includegraphics[width=0.46\columnwidth]{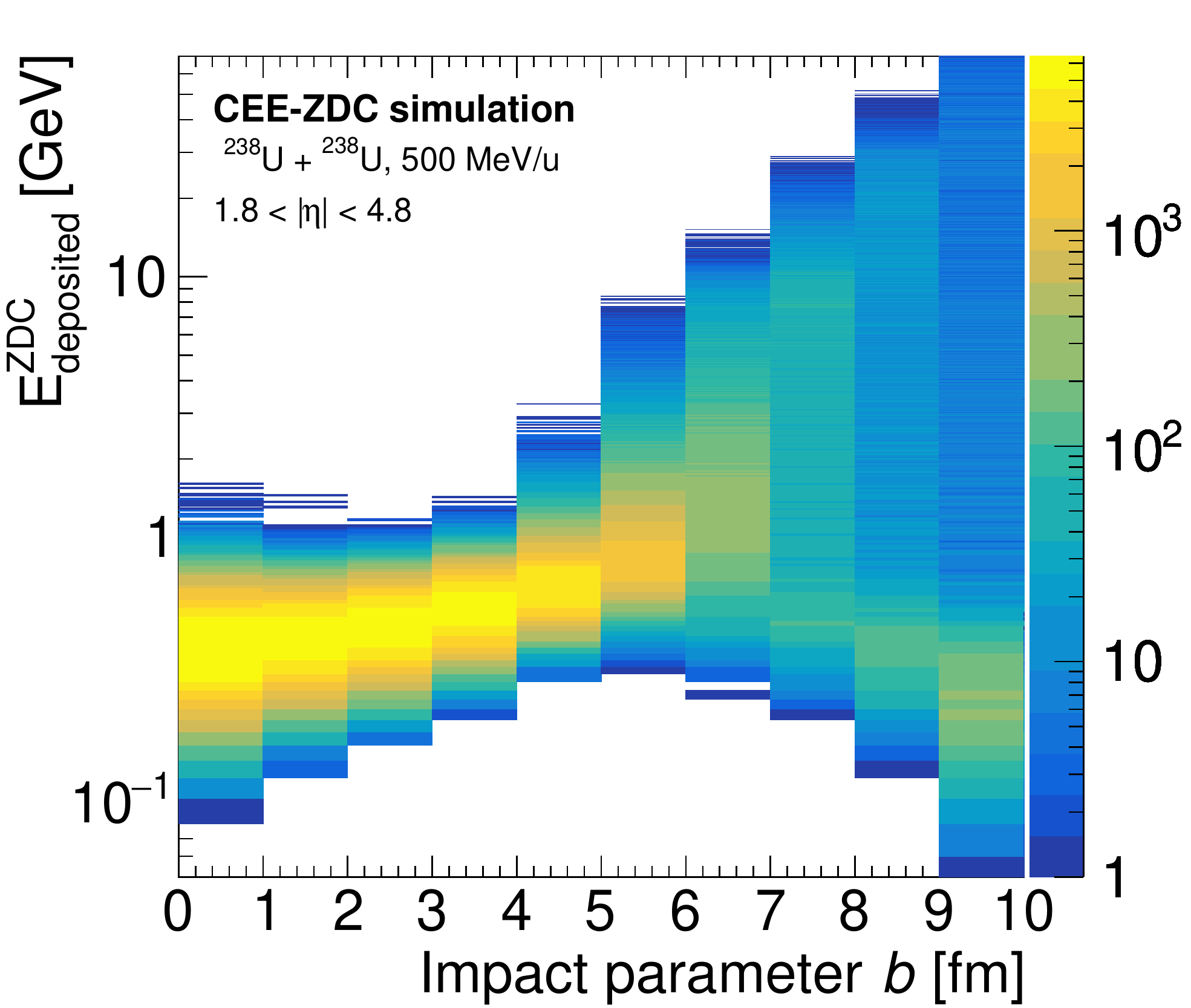}
  } 
  \caption{(a) The number of fired channels in ZDC as a function of impact parameters. (b)  The deposited energy in ZDC as a function of impact parameters.}     
  \label{Fig: monotonic_ZDC}
\end{figure}

The simulated data were generated by simulating $\rm ^{238}U$ + $\rm ^{238}U$ collisions at 500 MeV/$u$ with the IQMD generator~\cite{Hartnack:1997ez}, and the generated particles were then transported through the apparatus using the GEANT4 package~\cite{Brun:1994aa}. Determining centrality with only one forward-rapidity detector, like the ZDC, is challenging even when employing ML algorithms. Previous ML-based studies of centrality determination have relied on information from multiple subsystems within the detector, such as the tracks reconstructed from the central barrel detectors and deposited energy in forward calorimeters, revealing a strong correlation between the centrality class and observables. The CEE-ZDC is a non-tracking detector, and the number of spectator nucleons in a nucleus-nucleus collision is expected to be proportional to the deposited energy and the number of fired channels in the ZDC. However, the presence of a beam hole at the center of ZDC and limited detector acceptance result in a weak monotonic dependence between impact parameters and observables, as clearly illustrated in Fig.~\ref{Fig:2a} for the number of fired channels and Fig.~\ref{Fig:2b} for the deposited energy in the ZDC.

\begin{figure}[htb]
  \subfigure[] {
   \label{Fig:3a}     
  \includegraphics[width=0.46\columnwidth]{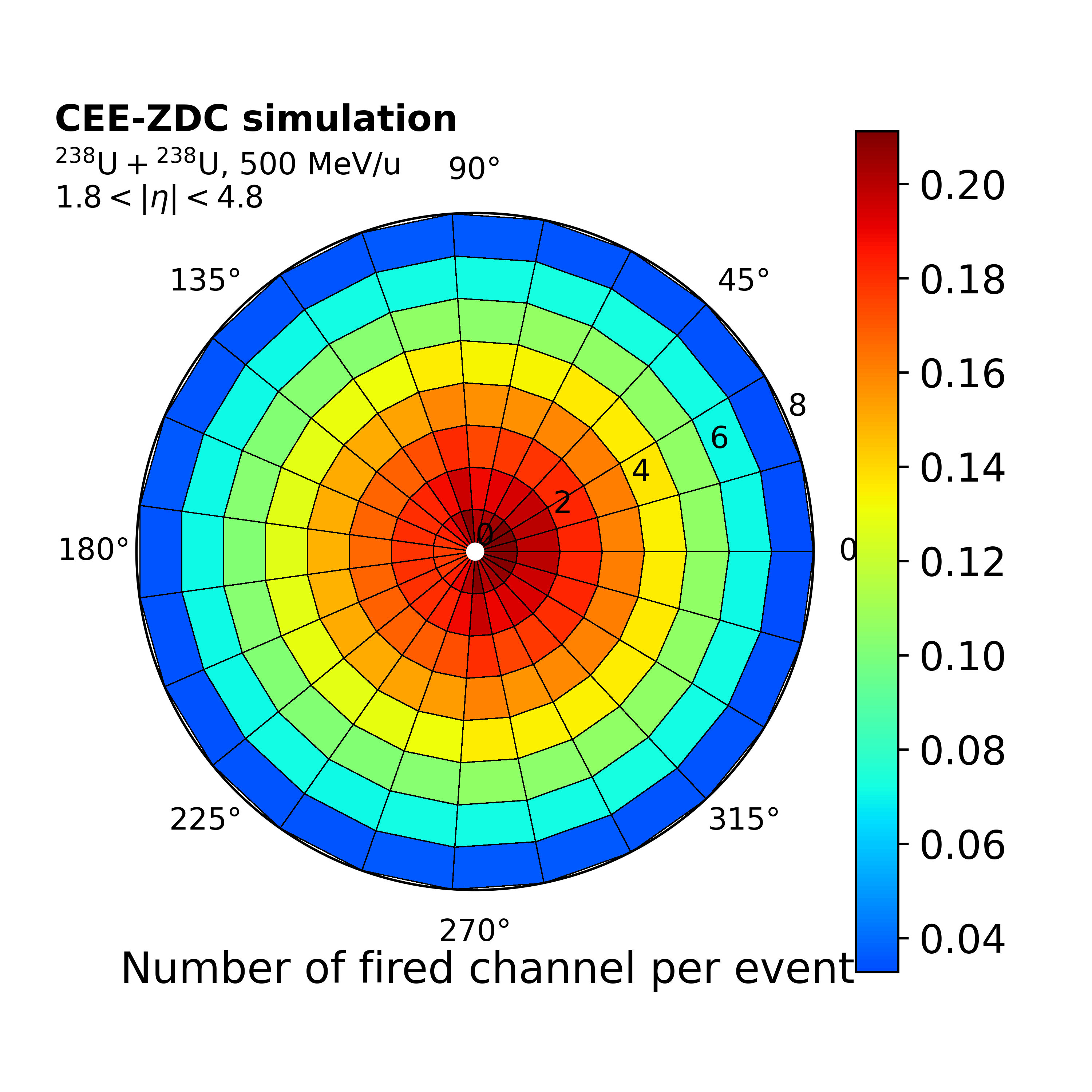}
  }
  \subfigure[] { 
  \label{Fig:3b}     
  \includegraphics[width=0.46\columnwidth]{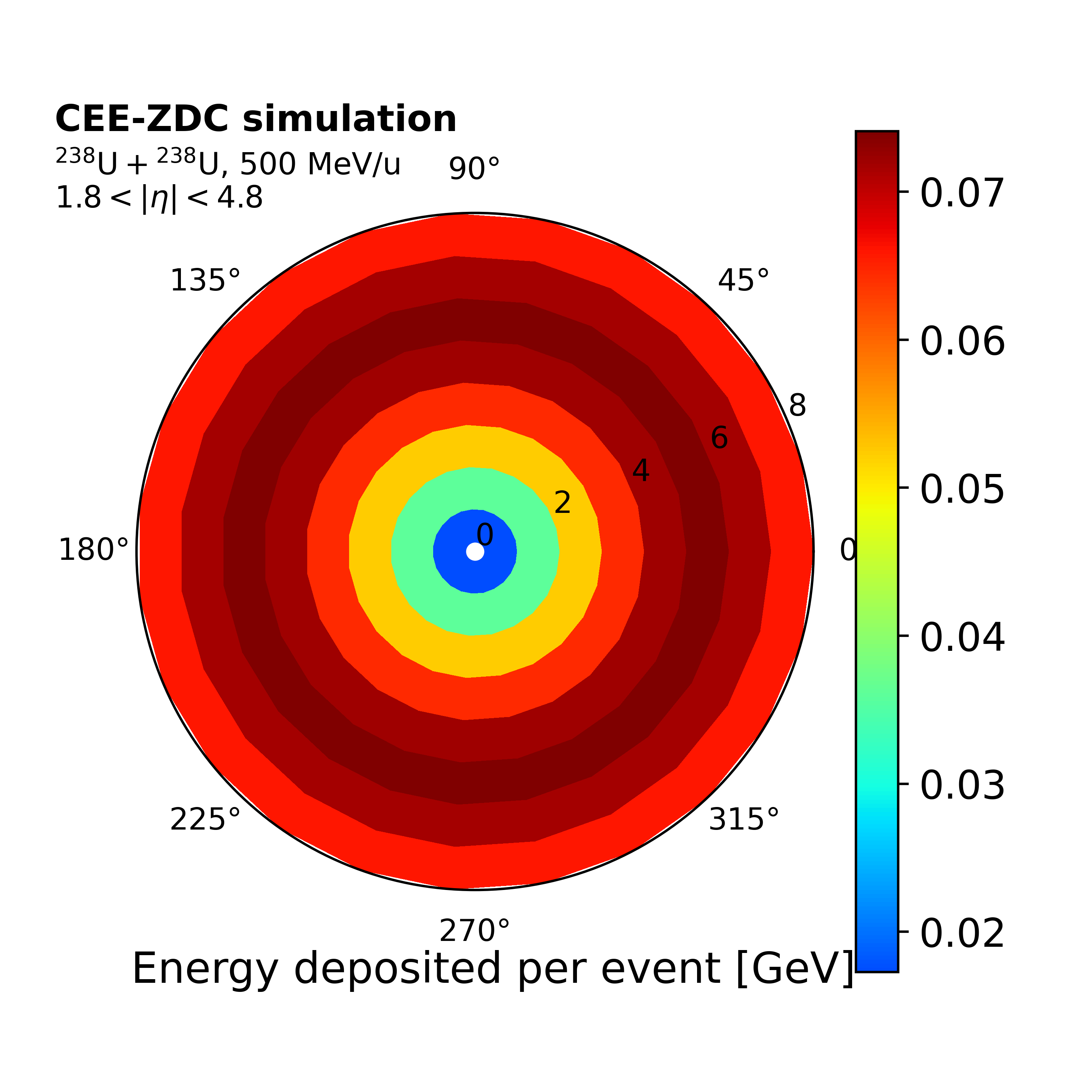}
  } 
  \caption{(a) Number of fired ZDC channel per event distribution in impact parameter interval of $7 < b \le 10 $ fm.
  (b) Deposited energy of ZDC ring per event distribution in impact parameter interval of $0 < b \le 3 $ fm.}    
  \label{Fig: Hit_distribution_2D}
\end{figure}

Potential improvements in centrality determination can be achieved by utilizing data from ZDC-subrings in conjunction with the ZDC as the additional features in ML task. Moreover, it may be advantageous to use the energy deposited in the ZDC ring-by-ring and the number of fired channels event-by-event, and to exploit all inherent correlations between modules. Fig.~\ref{Fig:3a} displays the number of fired ZDC channels per event distribution in the impact parameter range $7 < b \le 10$ fm, as well as the deposited energy per event distribution in the ZDC ring in the impact parameter range $0 < b \le 3$ fm shown in Fig.~\ref{Fig:3b}. The complex pattern and non-trivial decision boundary among the classes of event centrality present an ideal opportunity to apply ML techniques.

Boosted Decision Trees (BDT), as a family of popular supervised learning algorithms for classification and regression problems, are extensively used to analyze data in high-energy-physics experiments. Extreme gradient Boosting (XGboost), one of the powerful BDTs based on the gradient boosting method, is adopted to solve the multi-classification problems for the centrality determination in this work. The physics features used as the inputs for the model training are deposited energy in full ZDC and ZDC substrings as well as the number of fired channels in ZDC. The simulated data is divided into 3 centrality classes based on the impact parameter listed in Table.~\ref{tab: central_classes}. The samples are split into training and test samples of equal size for each centrality class. A state-of-the-art machine learning hyperparameter optimization with Optuna is adopted to speed up optimization time and achieve the best performance of the training models~\cite{akiba2019optuna}.

\begin{table}[ht]
  \centering
  \caption{The centrality classes with respect to the impact parameter $b$ intervals}
  \begin{tabular}{|c|c|}
    \hline
   Centrality class & {$b$ interval} [$\rm fm$] \\
    \hline 
    Central & $0 \le b \le 3 $    \\
    Semi-Central & $3 < b \le 7 $    \\
    Peripheral & $7 < b \le 10 $   \\

    \hline
  \end{tabular}
  \label{tab: central_classes}
\end{table}

\section{Performance of the ML models}\label{sec.IV}

\begin{figure}[htb]
  \subfigure[] {
   \label{Fig:4a}     
  \includegraphics[width=0.46\columnwidth]{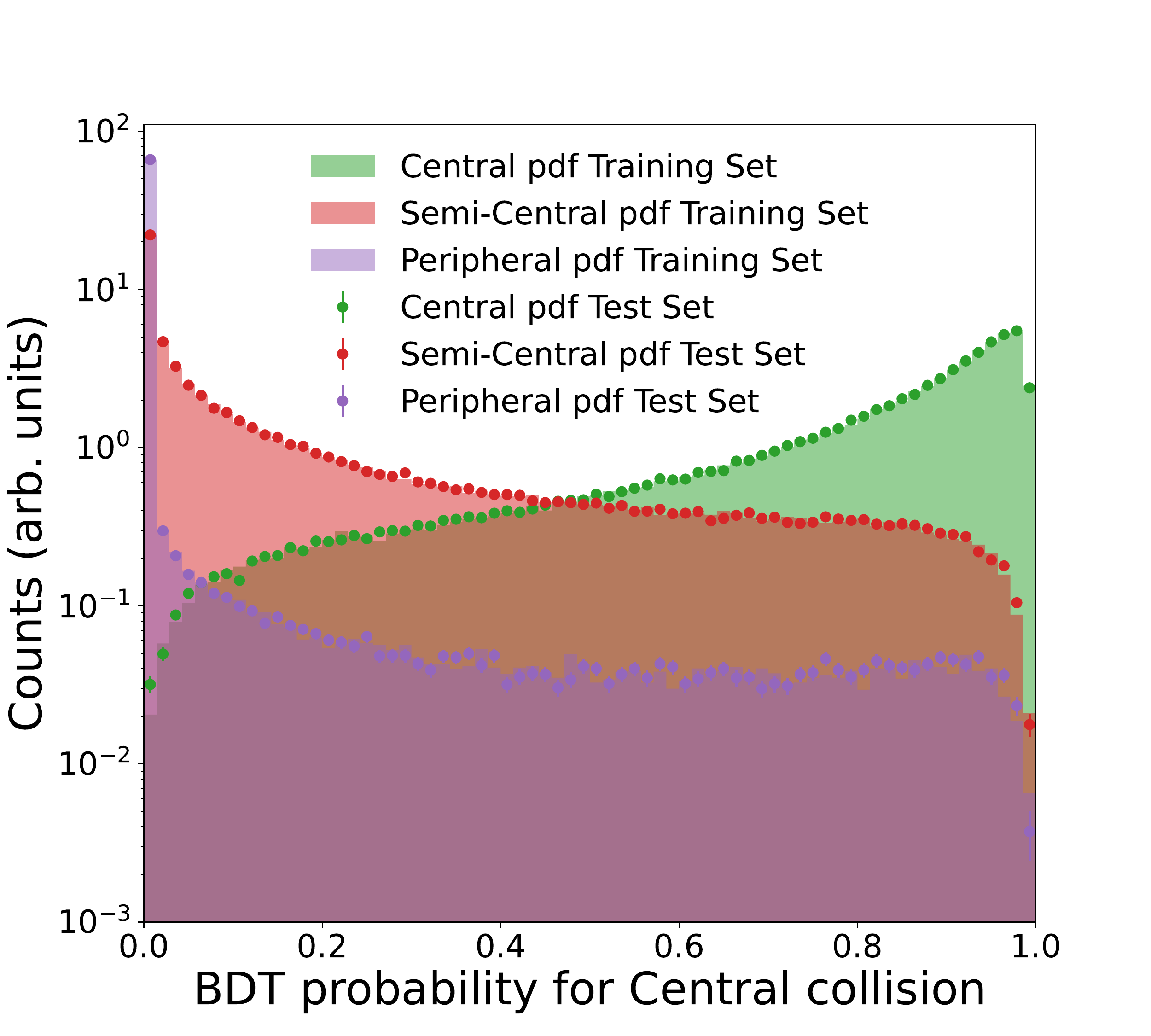}
  }
  \subfigure[] { 
  \label{Fig:4b}     
  \includegraphics[width=0.46\columnwidth]{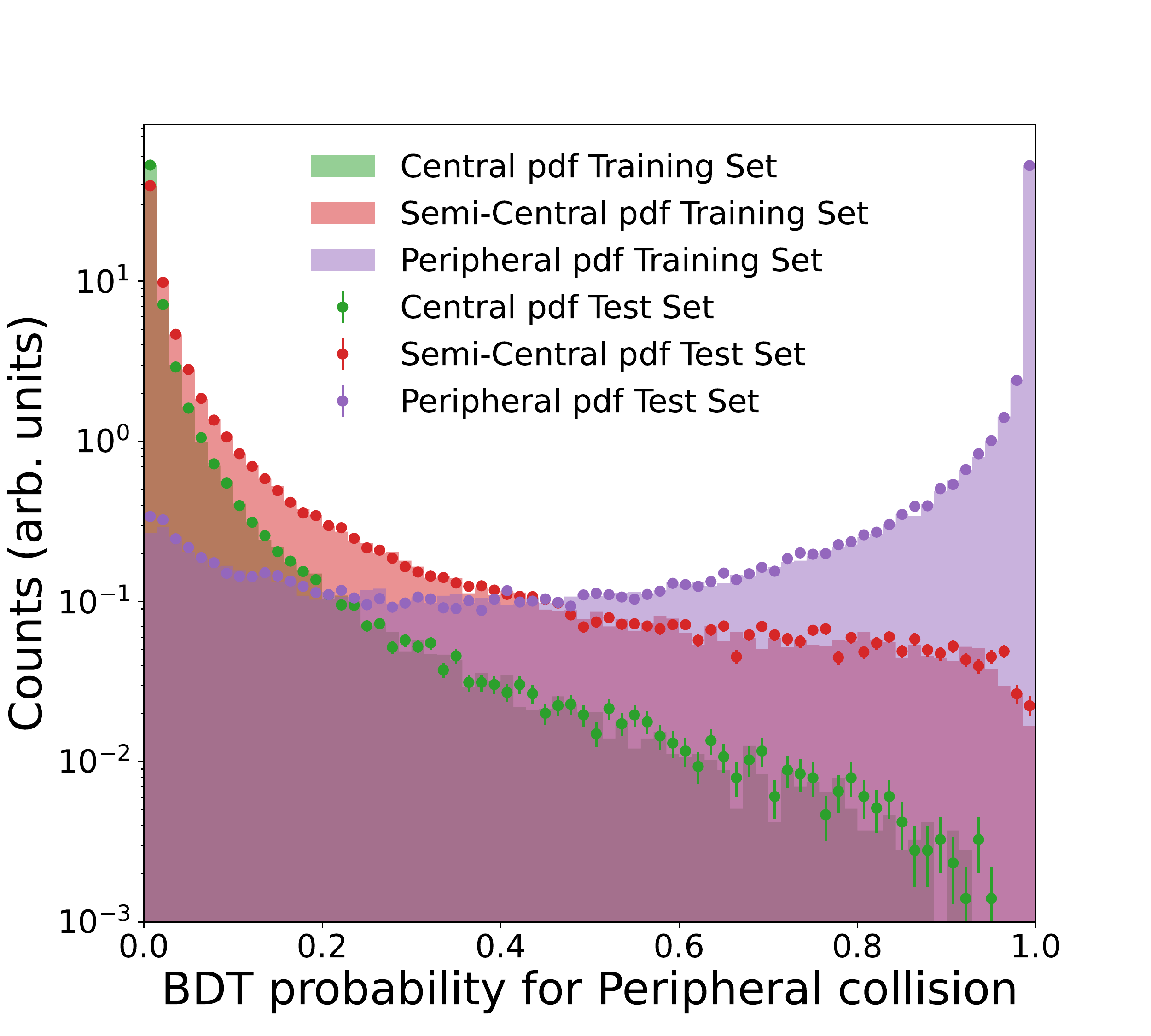}
  } 
  \caption{The probability distributions of belonging to the central class (a) and peripheral class (b) for both the training and test sets.}    
  \label{Fig: prob_dis}
\end{figure}

The machine learning model was applied to both the training and test sets to visualize the distributions of the ML output scores and check for consistency between the two sets. For classification with three centrality classes ($p_i$), the model generated three scores representing the probability of belonging to each of the considered classes. As per construction, the probabilities for each centrality class are summed to one ($\sum_{i=1}^3p_{i} = 1$). Fig.~\ref{Fig: prob_dis} illustrated the probability distributions of the central class (a) and peripheral class (b) for both the training and test sets. The probability distributions were close to unity for each probability distribution corresponding to the respective true class, while the other two distributions were shifted toward zero. The probability density function of the training and test samples for each centrality class showed good agreement, indicating that the model was not overfitting.

\begin{figure}[htbp]
  \centering   
  \includegraphics[width=0.8\columnwidth]{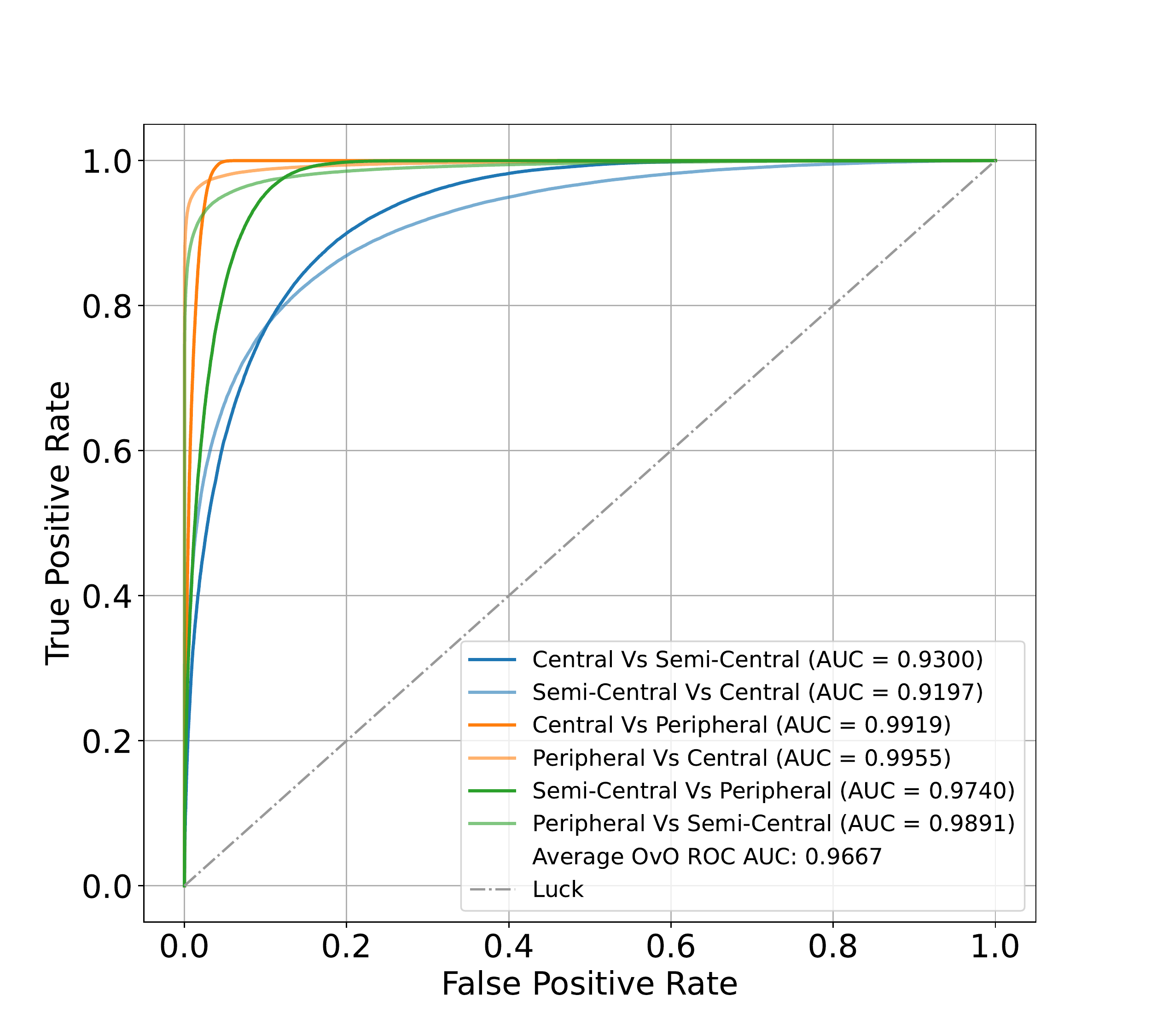}   
  \caption{ROC curves and AUCs with respect to different "One-vs-One" cases are shown with the different line colors.}
  \label{Fig: ROC}
\end{figure}

The Receiver Operating Characteristic (ROC) curve is commonly used to evaluate the performance of a classification model, by plotting the True Positive Rate against the False Positive Rate for various threshold settings. The area under the ROC curve, known as ROC AUC, provides a global measure of the model's performance, ranging from 0.5 (random classification) to 1 (perfect classification), independent of the threshold and class distribution~\cite{powers2020evaluation}. However, for multi-class classification, the ROC curve cannot be directly defined, and the "One-vs-One" approach is used to compute the overall average of the individual ROC AUCs for each pair of classes. In this study, the ROC curves and ROC AUC values obtained on the test set are reported in Figure~\ref{Fig: ROC}. The high final ROC AUC value of approximately 0.96 indicates that the BDT model is highly effective in determining centrality.

\section{Efficiency and purity of the centrality classification}\label{sec.V}

\begin{figure}[htbp]
  \centering   
  \includegraphics[width=0.8\columnwidth]{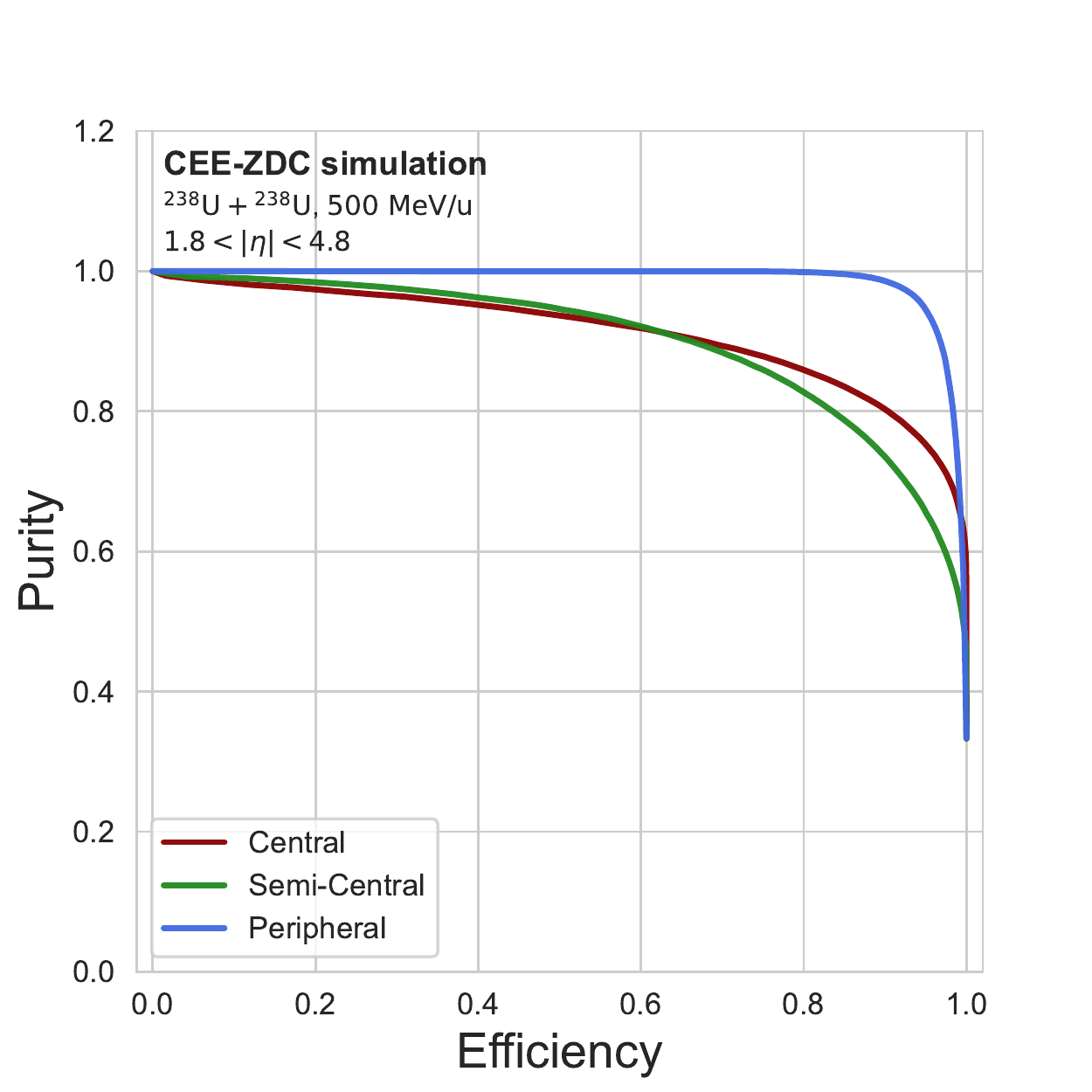}
  \caption{Efficiency versus purity of the multi-classification models for each centrality class. The red, green, and blue solid lines represent the central, semi-central, and peripheral classes, respectively.}     
  \label{Fig: eff_purity}
\end{figure}

The performance of the centrality classification model was evaluated by calculating the efficiency and purity based on the ML output scores. Efficiency refers to the fraction of correctly classified events, while purity measures the fraction of events correctly classified for a particular centrality class out of all the events assigned to that class. The efficiency versus purity of the multi-classification models for each centrality class is shown in Fig.~\ref{Fig: eff_purity}, where the red, green, and blue solid lines represent the central, semi-central, and peripheral classes, respectively. The peripheral class was found to be the most effectively classified, and the central class was found to be more challenging than the semi-central class in the higher efficiency region. The quantified values in Table.~\ref{tab: central_eff_purity} indicate that even at very high purity levels, the efficiency of the peripheral class is not significantly compromised, and both the central and semi-central classes also exhibit promising values for efficiency at the high purity. These results indicate that the ML-based event centrality determination utilized in ZDC is effective.

\begin{table}[ht]
  \centering
  \caption{Efficiency and purity values for different centrality classes}
  \begin{tabular}{|c|c|c|c|}
    \hline
    \diagbox{Purity}{Efficiency}{Class} & Central& Semi-Central &Peripheral\\
    \hline 
    Purity = 90\% & 67\% & 66\% & 97\%  \\
    Purity = 95\%  & 41\% & 47\% & 94\%  \\
    Purity = 98\%  & 11\% & 24\% & 93\%  \\
    \hline
  \end{tabular}
  \label{tab: central_eff_purity}
\end{table}

\begin{figure}[htb]
    \centering
    \includegraphics[width=1\hsize]{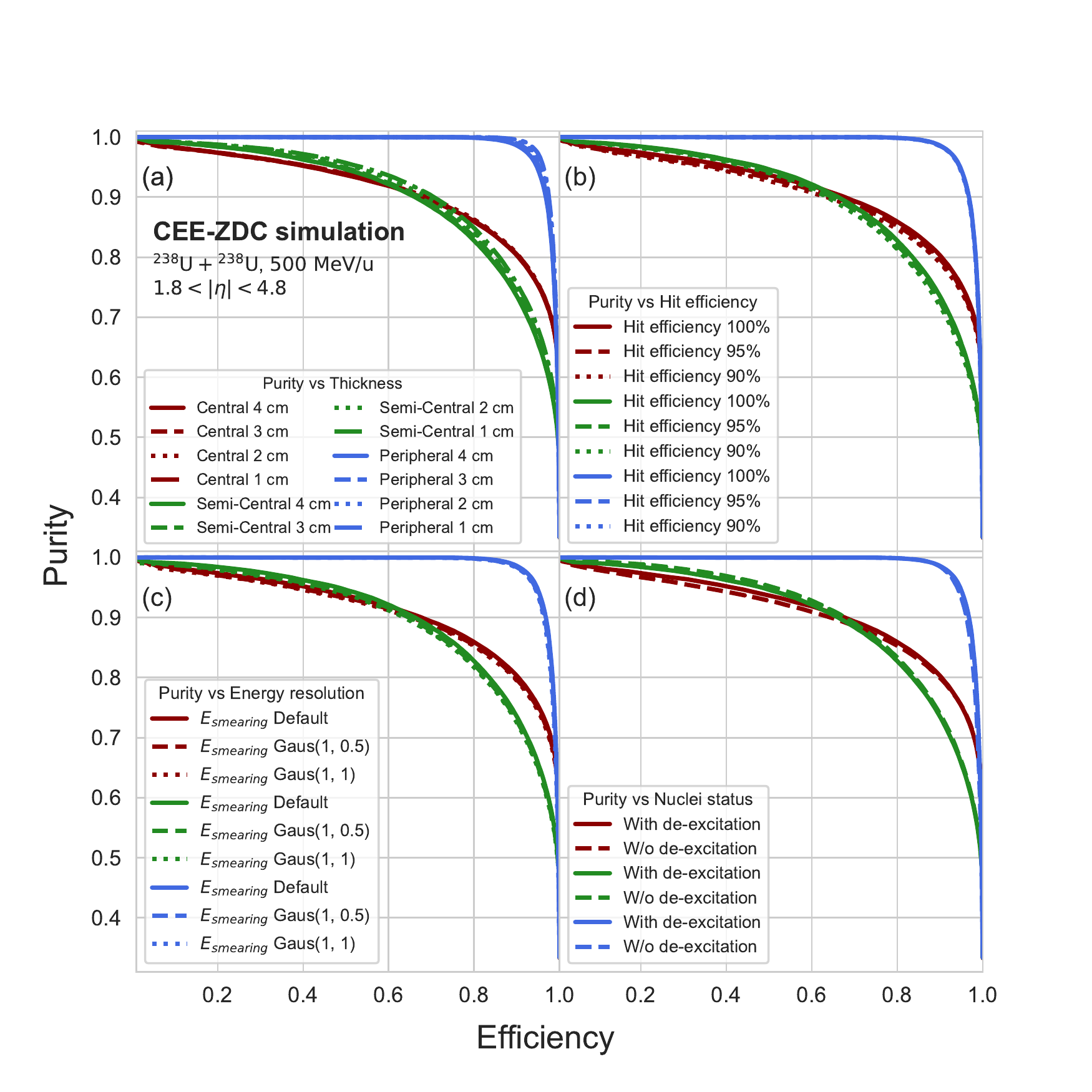}
    \caption{The effects of several factors on the efficiency and purity for the multi-classification models: (a) thickness of ZDC detector, (b) hit efficiency in ZDC (b), (c) energy resolution, (d) with or without de-excitation.}
    \label{Fig: effect_eff_pur}
  \end{figure}

In addition, to evaluate the performance of the centrality determination with ZDC, the effects of several factors related to the configuration of ZDC in the simulation data were systematically investigated. These factors included the thickness of the ZDC detector, hit efficiency, energy resolution, and heavy nuclei with de-excitation or without de-excitation (in the IQMD). The ZDC plastic scintillator thickness was varied from 1 cm to 4 cm, and the hit efficiency was varied with 90\% and 95\%. The energy deposited was also smeared with different sigma values of the Gaussian distributions. As shown in Fig.~\ref{Fig: effect_eff_pur}, the results indicated that the effect of these factors on the purity and efficiency of the centrality classification is minor. Among the tested factors, the ZDC detector thickness had the most significant impact, but even that effect is relatively small. In conclusion, it study suggests that the multi-classification adopted in ZDC is robust against variations in these factors, indicating the potential for reliable and accurate classification for the centrality with ZDC.

\section{Summary} \label{sec.VI}
The study aimed to determine the centrality class in nucleus-nucleus collisions at the CEE-ZDC detector using a multi-classification model based on the XGBoost classifier. The ML model was trained and tested from simulation data from the IQMD event generator and then modeled through the GEANT4 package. The additional study examined various factors associated with the geometry and response of the ZDC detector, and the results indicate that the impact of these factors is minor, demonstrating the robustness of the XGBoost classifier in determining centrality. Future work may include improving the accuracy of centrality determination by incorporating regression tasks and exploring other machine learning algorithms. The study indicates the good performance of the CEE-ZDC for centrality determination in nucleus-nucleus collisions.

\section*{Acknowledgments}
We thank Prof. Li Ou and Zhigang Xiao for generating IQMD data and fruitful discussions.

\bibliographystyle{unsrt}
\bibliography{reference}

\end{document}